\documentclass{article}

\usepackage{enumerate}
\usepackage{amsfonts}
\usepackage{amsmath}
\usepackage{comment}
\usepackage{textcomp}
\usepackage{amsthm, tikz}
\usepackage{amssymb}
\usepackage{color}
\usepackage{graphicx}
\usepackage{bbm}
\usepackage[matrix,arrow,curve]{xy}
\usepackage{array}
\usepackage{mathtools}

\def\be{\begin{eqnarray}}
\def\ee{\end{eqnarray}}
\def\nn{\nonumber}

\hoffset= - 1.3in         

\begin{document}

\title{{\bf {Colored knot polynomials for Pretzel knots and links\\of arbitrary genus
}\vspace{.2cm}}
\author{{\bf D. Galakhov$^{a,b,}$}\footnote{galakhov@itep.ru; galakhov@physics.rutgers.edu}, \ {\bf D. Melnikov$^{a,c,}$}\footnote{dmitry@iip.ufrn.br}, \  {\bf A. Mironov$^{d,a,e,}$}\footnote{mironov@lpi.ru; mironov@itep.ru}, \ {\bf A. Morozov$^{a,e,}$}\thanks{morozov@itep.ru} \ and {\bf
A. Sleptsov$^{a,e,f,}$}\footnote{sleptsov@itep.ru}}
\date{ }
}

\maketitle

\vspace{-6.0cm}

\begin{center}
\hfill FIAN/TD-19/14\\
\hfill ITEP/TH-42/14\\
\end{center}

\vspace{4.2cm}

\begin{center}
$^a$ {\small {\it ITEP, Moscow 117218, Russia}}\\
$^b$ {\small {\it NHETC and Department of Physics and Astronomy, Rutgers University,
Piscataway, NJ 08855-0849, USA }}\\
$^c$ {\small {\it International Institute of Physics, UFRN
Av. Odilon G. de Lima 1722, Natal 59078-400, Brazil}}\\
$^d$ {\small {\it Lebedev Physics Institute, Moscow 119991, Russia}}\\
$^e$ {\small {\it National Research Nuclear University MEPhI, Moscow 115409, Russia }}\\
$^f$ {\small {\it Laboratory of Quantum Topology,
Chelyabinsk State University, Chelyabinsk 454001, Russia}}
\end{center}

\vspace{1cm}

\begin{abstract}
A very simple expression is conjectured for arbitrary colored
Jones and HOMFLY polynomials of a rich $(g+1)$-parametric family of
Pretzel knots and links.
The answer for the Jones and HOMFLY
is fully and explicitly expressed through the Racah
matrix of $U_q(SU_N)$,
and looks related to
a modular transformation of toric conformal block.
\end{abstract}

\vspace{2cm}

Knot polynomials \cite{knotpols} are among the hottest topics in modern theory.
They are supposed to summarize nicely representation theory of
quantum algebras and modular properties of conformal blocks \cite{Wit}-\cite{Gal}.
The result reported in the present letter, provides a spectacular
illustration and support to this general expectation.

\bigskip

The genus-$g$  knot/link, also known as Pretzel \cite{Pre}, is shown in the picture:

\begin{figure}[h!]
\centering\leavevmode
\includegraphics[width=18 cm]{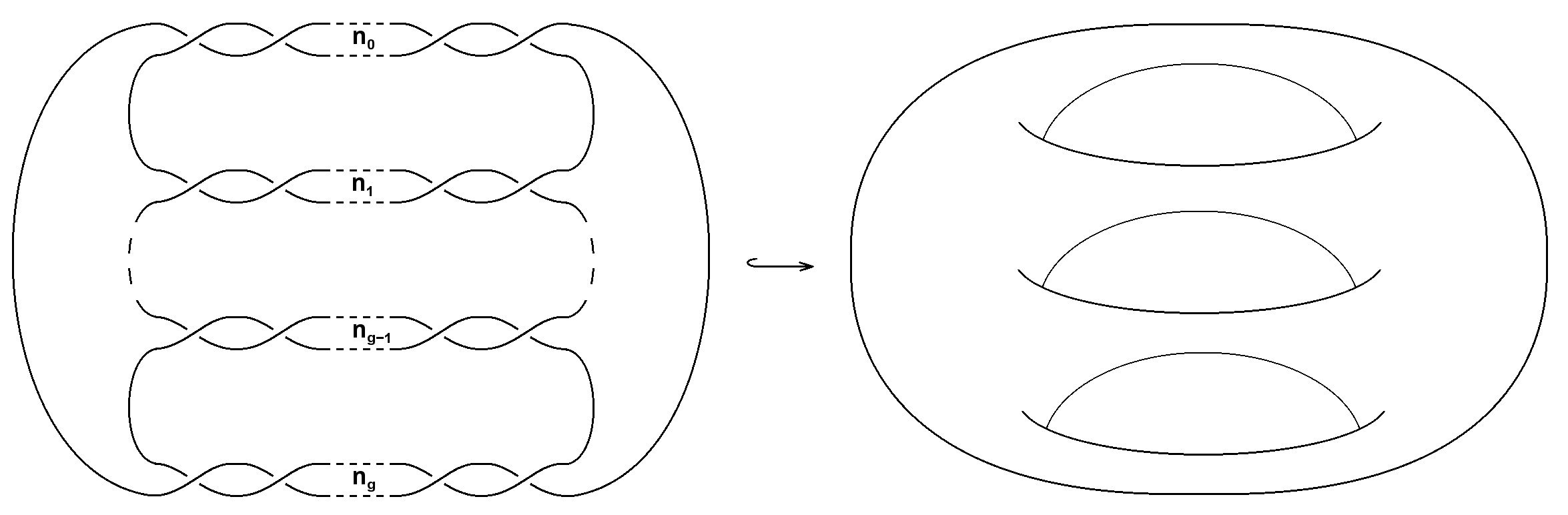}
\caption{Genus $g$ knot}
\label{braidg}
\end{figure}

\noindent
It depends on $g+1$ integers $n_0,\ldots,n_g$, the algebraic lengths
of the constituent $2$-strand braids.
Orientation of lines does not matter, when one considers the Jones polynomials
(not HOMFLY!).
Also, these polynomials are defined only for the symmetric representations $[r]$
and, hence, do not change under arbitrary permutations of parameters $n_i$
(though the knot/link itself has at best the cyclic symmetry
$n_i\longrightarrow n_{i+1}$, and even this is true only for particular orientations).
A particular manifestation of this enhanced symmetry has been recently
noted in \cite{ensym}.

\bigskip

The answer for the colored Jones polynomials for this entire family can be
written in full generality, and is wonderfully simple
\be
\boxed{
J_r^{(n_0,\ldots, n_g)}(q) = \sum_{k=0}^r \ [2k+1] \
\prod_{i=0}^g \left(\sum_{m=0}^r {\cal A}_{km}\lambda_m^{n_i}\right)
}
\label{Jans}
\ee
where
\be
\lambda_m = (-)^m q^{m(m+1)}
\ee
and
\be
{\cal A}_{km} = \sqrt{\frac{[2m+1]}{[2k+1]}}\cdot
S_{km}\left(\begin{array}{cc}  r & r \\ r & r \end{array}\right)
\label{AS}
\ee
is made out of the Racah matrix \cite{Racah}
of $SU_q(2)$ in representation $[r]$ (spin $r/2$).
Orthogonality of $S$ implies that
\be
\sum_{m=0}^r \frac{{\cal A}_{km}{\cal A}_{k'm}}{[2m+1]} = \frac{\delta_{k,k'}}{[2k+1]},\nn \\
\sum_{k=0}^r [2k+1]\, {\cal A}_{km}{\cal A}_{km'} = [2m+1]\delta_{m,m'}
\ee
Quantum numbers in these formulas are defined as $[n]=\frac{q^n-q^{-n}}{q-q^{-1}}$.

\bigskip

The first of the $r+1$ polynomials
\be
P_k^{(n)}(q|r) = \sum_{m=0}^r {\cal A}_{km}\, \lambda_m^n
\ee
is just the Jones polynomial for the 2-strand torus knot/link $T[2,n]$:
\be
P_0^{(n)}(q|r) = J^{(n)}_r = \sum_{m=0}^r [2m+1] \cdot \lambda_m^n
\ee
The origin of its orthonormal "satellites" $P_k^{(n)}(q|r)$ with $k=1,\ldots,r$
and of entire rotation from the monomial basis $\{\lambda_m^n\}$
is yet unknown.

\bigskip

Naturally, a straightforward  generalization exists
to the HOMFLY polynomials:
\be
\boxed{
H_{\vec R}^{\vec n}(A,q) = \sum_{X\, \in\, \bigcap_i R_i\otimes \bar R_i}  {\rm dim}_{_X}
\cdot \prod_{i=0}^{g} \left( \sum_{Y_i\in R_i\otimes R_{i+1}}
{\cal A}_{XY_i}\left(\begin{array}{cc} \bar R_i & \bar R_{i+1}\\ R_i & R_{i+1} \end{array}\right)
\cdot \lambda_Y^{n_i}\right)
}
\label{Hans}
\ee
which looks  like a modular transformation of {\it toric} conformal block,
summed over intermediate states $X$ in the loop:

\begin{picture}(100,110)(-200,-50)
\put(0,0){\circle{40}}
\put(-19,-5){\line(-1,0){20}}  \put(-30,-5){\vector(1,0){2}}
\put(-19,5){\line(-1,0){20}}   \put(-31,5){\vector(-1,0){2}}
\put(19,-5){\line(1,0){20}}     \put(30,5){\vector(-1,0){2}}
\put(19,5){\line(1,0){20}}      \put(31,-5){\vector(1,0){2}}
\put(-6,19){\line(0,1){21}}      \put(-6,30){\vector(0,1){2}}
\put(6,19){\line(0,1){20}}      \put(6,32){\vector(0,-1){2}}
\put(-6,-19){\line(0,-1){20}}    \put(-6,-32){\vector(0,1){2}}
\put(6,-19){\line(0,-1){20}}     \put(6,-30){\vector(0,-1){2}}
\put(0,20){\vector(-1,0){2}}
\put(20,0){\vector(0,-1){2}}
\put(0,-20){\vector(1,0){2}}
\put(-20,0){\vector(0,1){2}}
\put(-40,12){\mbox{  {\footnotesize $R_i$}  }}
\put(-17,-4){\mbox{  {\footnotesize $R_i$}  }}
\put(-31,35){\mbox{  {\footnotesize $R_{i+1}$}  }}
\put(7,35){\mbox{   {\footnotesize $R_{i+1}$}  }}
\put(-8,8){\mbox{    {\footnotesize $R_{i+1}$}  }}
\put(15,17){\mbox{$X$}}
\put(15,-25){\mbox{$X$}}
\put(-21,-25){\mbox{$X$}}
\put(-21,17){\mbox{$X$}}
\qbezier(-6,40)(-6,55)(-55,40)
\qbezier(-39,5)(-50,5)(-55,40)
\qbezier(-55,40)(-57,50)(-78,44)
\qbezier(-55,40)(-67,35)(-75,49)
\put(-70,30){\mbox{$n_i$}}
\end{picture}

\noindent
Here ${\rm dim}_{_X}$ and ${\cal A}$ are the universal $A$-dependent dimension of representation $X$ (which is equal at 
$A=q^N$ to the quantum dimension of representation $X$ of $SU_q(N)$)
and rescaled Racah matrix respectively, interpolating between
those for $SU_q(N)$ at $A=q^N$.
This formula is indeed true \cite{Sle}, at least when all the representations
are symmetric or their conjugate: $R_i=[r],\overline{ [r]}$.
The answer for antisymmetric representations then follows from the general transposition rule
\cite{DMMSS,GS}
$H_{[1^r]} (A,q) = H_{[r]}(A,q^{-1})$.
It should possess further continuation
{\it a la} \cite{IMMMfe} to superpolynomials, thus providing a
$\beta$-deformation \cite{betadefo} of the universal Racah matrix.

\bigskip

Eqs.(\ref{Jans}) and (\ref{Hans}) result from a tedious calculation
in \cite{Sle,GalS} with the help of
evolution \cite{DMMSS,evo}, modernized Reshetikhin-Turaev
\cite{RT,RTmod}
and modular matrix \cite{Wit,inds,Gal} methods.
Of course, a simple conceptual derivation should exist for such a simple
and general formula, but it still remains to be found.
The value of (\ref{Jans}) is independent of the derivation details,
and it is high, because this only formula (together with its lifting to the HOMFLY polynomials
in \cite{Sle}) contains almost all
what is currently known about explicit colored knot polynomials
beyond torus links: in particular, all the twist and 2-bridge knots are small subsets
in the Pretzel family
(however, among torus knots with more than two strands, only $[3,4]$ and $[3,5]$ belong to it).

\bigskip

\paragraph{Examples:}

We give them here mostly for the Jones case, for an  exhaustive description
of the symmetric HOMFLY polynomials for all Pretzel links see \cite{Sle}.
First of all, we list the first few $S$ and ${\cal A}$ matrices
for the lowest representations of $SU_q(2)$:

\be
r=1:     & S = \frac{1}{[2]}\left(\begin{array}{cc}
 1 & \sqrt{[3]} \\ \sqrt{[3]} & -1
\end{array} \right)
\ \ \ \  &  {\cal A}  = \frac{1}{[2]}\left(\begin{array}{cc}
 1 & [3] \\ 1 & -1
\end{array} \right),
\nn \\ \nn \\ \nn \\
r=2: & S= \frac{1}{[3]}\left(\begin{array}{ccc} 1 & \sqrt{[3]} & \sqrt{[5]} \\ \\
\sqrt{[3]} & \frac{[6]}{[4]} & -\frac{[2]\sqrt{[3][5]}}{[4]} \\ \\
\sqrt{[5]}& -\frac{[2]\sqrt{[3][5]}}{[4]} & \frac{[2]}{[4]}  \end{array}\right),
& {\cal A} =  \frac{1}{[3]}\left(\begin{array}{ccc} 1 & [3] & [5] \\ \\
1 & \frac{[6]}{[4]} & -\frac{[2][5]}{[4]} \\ \\
1& -\frac{[2][3]}{[4]} & \frac{[2]}{[4]}  \end{array}\right),
\nn \\ \nn \\ \nn \\
r=3: & S=  \frac{1}{[4]}\left(\begin{array}{cccc} 1 & \sqrt{[3]} & \sqrt{[5]} & \sqrt{[7]} \\ \\
\sqrt{[3]} & \frac{[2][6]-1}{[5]}  & \Big([5]-[2]^2\Big)\sqrt{\frac{[3]}{[5]}} &
-\frac{[3]^2[7]}{\sqrt{[3][5]}}\\ \\
\sqrt{[5]}& \frac{[7]-[2]^2}{\sqrt{[3][5]}}  & -\frac{[2]\,\big([2][6]-[3]\big)}{[6]}
&  \frac{[2][3]}{[6]}\sqrt{\frac{[7]}{[5]}} \\ \\
\sqrt{[7]} & -\frac{[3]}{[5]}\sqrt{[3][7]} & \frac{[2][3]}{[6]}\sqrt{\frac{[7]}{[5]}}
&  -\frac{[2][3]}{[5][6]}
\end{array}\right),
& {\cal A} = \frac{1}{[4]} \left(\begin{array}{cccc} 1 & [3] & [5] & [7] \\ \\
1 & \frac{[2][6]-1}{[5]}  & [5]-[2]^2 &  -\frac{[3][7]}{[5]}\\ \\
1& \frac{[7]-[2]^2}{[5]}  & -\frac{[2]\,\big([2][6]-[3]\big)}{[6]}
&  \frac{[2][3][7]}{[5][6]} \\ \\
1 & -\frac{[3]^2}{[5]} & \frac{[2][3]}{[6]}  &  -\frac{[2][3]}{[5][6]}
\end{array}\right), \nn \\ \nn \\
& \ldots \nn \\
\nn
\ee
In general, from (\ref{AS}) and \cite{Racah} one has:
\be
{\cal A}_{km}=(-1)^{r+k+m} [2m+1]\cdot{\Big([k]!\,[m]!\Big)^2\,[r-k]!\,[r-m] !\over [r+k+1]!\,[r+m+1]!}
\ \sum_j {(-1)^j\,[j+1]!\over \Big([j-r-k]!\,[j-r-m]!\,[r+k+m-j]!\Big)^2[2r-j]!}
\ee
Given these matrices, eq.(\ref{Jans}) provides absolutely explicit
expressions for {\it all} genus-$g$  knots/links in the corresponding representations
\cite{GalS}:
\be
J_1^{(n_0,\ldots, n_g)} =
\frac{1}{[2]^{g+1}}
\left\{
\prod_{i=0}^g  \Big(1+[3]\cdot (-q^2)^{n_i}\Big) +
[3]\cdot \prod_{i=0}^g \Big( 1 \, - \, (-q^2)^{n_i}\Big)\right\},
 \\ \nn \\ \nn \\
J_2^{(n_0,\ldots, n_g)} =
\frac{1}{[3]^{g+1}}
\left\{
\prod_{i=0}^g   \Big(1+ [3]\cdot(-q^2)^{n_i} + [5]\cdot q^{6n_i} \Big)
\ + \right. \nn \\ \left. + \
[3]\cdot\prod_{i=0}^g
\left(1 + \frac{[6]}{[4]}\cdot (-q^2)^{n_i}-\frac{[2][5]}{[4]}\cdot q^{6n_i}\right)+
[5]\prod_{i=0}^g
\left(1 -\frac{[2][3]}{[4]}\cdot (-q^2)^{n_i}+\frac{[2]}{[4]}\cdot q^{6n_i}\right)\right\},
\label{jones2} \\ \nn \\
\ldots \nn
\ee
Similarly, the fundamental  HOMFLY from \cite{Sle} is
\be
H_{[1]}^{(n_0,\ldots, n_g)} =
\frac{1}{\chi_{[1]}^{g+1}}
\left\{
\prod_{a=0}^{g-2g_{||}}\Big(1+\Delta_1 \cdot (-A)^{n_a}\Big)
\prod_{i=1}^{2g_{||}} \Big(\chi_{[11]} + \chi_{[2]}\cdot (-q^2)^{n_i}\Big)\  +
\right. \nn \\ \left. + \
\Delta_1  \prod_{a=0}^{g-2g_{||}}\Big(1 -  (-A)^{n_a}\Big)
\prod_{i=1}^{2g_{||}} \left(\frac{\chi_{[1]}}{[2]} \Big(1\,-\,(-q^2)^{n_i}\Big)\right)\right\}
\ee
where $\chi_{m}$ and $\Delta_k$ are quantum dimensions of the representations appearing in
the products $[r]\otimes [r]$ and $[r]\otimes \overline{[r]}$ respectively
(i.e. restrictions of the Schur functions to the topological locus, described by the hook formulas).
One more new parameter is the number $2g_{||}$ of parallel braids
($g_{||}=g_{\uparrow\uparrow}=g_{\downarrow\downarrow}$),
the remaining $g_{\uparrow\downarrow} = g+1-2g_{||}$ are antiparallel.

\bigskip

As to practical applications, formula (\ref{jones2}) for, say,
\be
J_{[2]}^{(\overbrace{1,\ldots, 1}^a ,\overbrace{n,\ldots,n}^b)} =
\frac{1}{[3]^b}\left\{q^{8a}\cdot\Big(1\mp q^{2n}[3]+q^{6n}[5]\Big)^b+
\right. \nn \\ \left. +
[3]\cdot q^{6a}
\left(1-q^{6n}-(q^4+1+q^{-4})\cdot q^{2n+2}\cdot\frac{q^{4n}\pm 1}{q^4+1}\right)^b
+ [5]\cdot q^{2a} \left(1\pm q^{2n}+q^{2n+2}\cdot\frac{q^{4n}\pm 1}{q^4+1}\right)^b\right\}
\ee
(upper/lower signs are for odd/even $n$)
can look a little too long as compared to the usual
$$ \ 1 \mp [3]\cdot q^{2n+1} + [5]\cdot q^{6n+6}\ $$  for the
two-strand knots/links $[2,n+1]=(n+1)_1$ (when $a=1,b=1$) or
 $$\ [3]-  [3][4][n+1]\{q\}^2\cdot q^{n+3}  + {[2][5]}[n+1]\{q\}^4\cdot q^{2n+6}\,
\left([n+2]+q^{n+3}[2n+1]\right),\ \ \ \ \ \ \ \  \{q\}=q-q^{-1}\ $$
for the twist knots $(n+2)_2$ or $(|n|+1)_1$ (when $a=2,b=1, \ n$ odd):
a lot of cancelations happen in
these special cases.
However, in generic situation such cancelations do not take place, and
just the same expression describes the colored Jones polynomials
for, say, the Pretzel $(1,1,1,1,1,1,1,1,19,19,19,19,19,19,19)$ with $a=8,b=7,n=19$,
which is about one page long.

To avoid possible confusion, we note that the Jones polynomials are unreduced
and appear in this formalism
in the vertical framing, the one consistent with the orientation
independence.
Conversion to the topological framing inserts the factor $q^{-8}$ for each vertex in
the parallel braids,
this means a total $q^{-8(n+1)}$ for $(n+1)_1$, while for the twist knots
the total factor is either $q^{16}$ or just $1$, if $tw_k$ is represented as $(-1,-1,2k)$
or as $(1,1,2k-1)$.


\bigskip

{\bf Acknowledgements.}
We are indebted for hospitality to the International Institute of Physics (Natal, Brazil),
where a significant part of this work was done. 

Our work is partly supported by grants NSh-1500.2014.2,
by RFBR grants 13-02-00457 (D.G., A.Mir. \& A.S.), 13-02-00478 (A.Mor.),  14-02-00627 (D.M.),
by the joint grants 13-02-91371-ST-a and 15-51-52031-NSC-a (A.Mir.,A.Mor. \& A.S.),
by 14-01-92691-Ind-a (D.M., A.Mir., A.Mor. \& A.S.)
and by the young-scientist grants  14-01-31395-young-a (D.G.) and 14-02-31446-young-a (A.S.).
Also we are partly supported by
the Brazilian National Counsel of Scientific and Technological Development (D.M. \& A.Mor.), by foundation FUNPEC-UFRN (A.Mir. \& A.S.),
by the Quantum Topology Lab of Chelyabinsk State University
(Russian Federation government grant 14.Z50.31.0020) (A.S.)
and by DOE grants SC0010008, ARRA-SC0003883, DE-SC0007897 (D.G.).

\end{document}